# Thermodynamic uncertainty relations for multi-terminal systems with broken time-reversal symmetry

Yanchao Zhang [1,*], Xinzhi Liu [1], Xiaolong Lü[1], Shanhe Su [2,†]

[1] *School of Science, Guangxi University of Science and Technology, Liuzhou 545006, People's Republic of China*

[2] *Department of Physics, Xiamen University, Xiamen 361005, People's Republic of China*

We investigate the thermodynamic uncertainty relations (TURs) in steady-state transport for a multi-terminal system consisting of two conducting terminals and *N*-2 probe terminals, within the linear response regime under broken time-reversal symmetry. We independently derive exact bounds on the TURs for the steady-state particle and heat currents under a strong constraint on the Onsager coefficients. Based on our proposed exact bounds, the analysis reveals that the bounds differ for particle and heat currents and are dependent on the system parameters. Furthermore, we demonstrate that under specific parameter conditions, the TURs of the particle and heat currents have a unified minimum value that depends solely on the number of terminals.

* Email: zhangyanchao@gxust.edu.cn

† Email: shanhesu@xmu.edu.cn

## I. INTRODUCTION

The thermodynamic uncertainty relation (TUR), which characterizes the trade-off between the entropy production rate (thermodynamic cost) and fluctuations (precision), has recently been established in nonequilibrium thermodynamics. This principle is a powerful quantitative tool for describing nonequilibrium physical systems [1-3]. The TUR was originally discovered in the nonequilibrium steady states of classical Markov processes [4] and was later rigorously proven using the large deviation theory [5, 6]. The fundamental principles and validity of the TUR have been extended beyond Markov processes [4, 7-10] to periodic driven processes [11, 12], from continuous-time [5, 6, 13, 14] to discrete-time [15, 16], and from classical systems [17-20] to quantum systems [21-24]. Experimental verifications of the TUR have also been performed [25, 26]. Furthermore, the TUR has been extended to various scenarios, including measurement and feedback processes [27]. It has also been used to bound the temporal extent of anomalous diffusion in finite systems driven out of equilibrium [28, 29]. Moreover, the TUR has been applied to describe current responses to kinetic perturbations [30, 31] and coherent transport phenomena [32, 33]. The TUR has inspired extensive research focused on understanding its origin and impact on heat engines (including quantum heat engines). Recent studies suggest that the fluctuation theorem can directly establish a bound known as the generalized TUR, which possesses broader applicability and greater universality [34-37]. For heat engines, the TUR establishes a stronger trade-off between power, efficiency, and stability, offering deeper insights into the second law of thermodynamics in nonequilibrium regimes [38-44]. Thus, the TUR may be seen as the natural counterpart of the fluctuation-dissipation theorem [34,45] or as a more precise formulation of the second law of thermodynamics [46].

The TUR can be quantified by the dimensionless parameter $Q_J$, which provides a fundamental limit on the precision of nonequilibrium currents in terms of the fluctuation and entropy production rate. The TUR for two-terminal systems in a nonequilibrium steady state is explicitly given by the following expression

$$Q_J \equiv \frac{D_J \sigma}{J^2 k_B} \geq 2, \tag{1}$$

where $J$ represents the steady-state current, $D_J$ denotes the current fluctuation, $\sigma$ is the entropy production rate, and $k_B$ is Boltzmann constant. This inequality provides a lower bound $Q_J \geq 2$ within the linear response regime [4]. The TUR clarifies the fundamental constraints between the fluctuations of thermodynamic quantities and the system's entropy production rate. This principle illustrates how energy dissipation limits microscopic fluctuations, with its bound representing the theoretical limit for optimizing system performance and stability. In certain cases, the bound can fall below 2, suggesting a potential violation of the TUR as expressed in Eq. (1). The possibility of such violation has been theoretically predicted in quantum dot junctions with noninteracting electrons [47, 48], nonequilibrium spin-boson systems and fermionic chains [49], periodically driven work-to-work converters [50], three-level masers [51], and classical pendulum clocks [52]. Additionally, the geometric properties of quantum nonequilibrium steady states inherently suggest the existence of a quantum TUR, i.e., $D_J \sigma / J^2 k_B \geq 1$, which is twice as loose as the classical bound given in Eq. (1) [53]. Further relaxation of the TUR bound in its original form has also been observed in systems with a magnetic field that breaks time-reversal symmetry [54-58]. Thus, understanding the underlying mechanisms behind these violations of the TUR and exploring the potential establishment of more general bounds remain key questions in ongoing research [51]. When the time-reversal symmetry is broken in the presence of a magnetic field, there is a stronger bound on the Onsager coefficients emerges for multi-terminal systems compared to two-terminal systems [61]. This strong bound on the Onsager coefficients has a stronger constraint than the second law of thermodynamics, and it is not clear how this will affect the TUR of multi-terminal systems. The objective of this paper is to we investigate the TUR of steady-state transport for multi-terminal systems within the linear response regime. We will independently derive exact bounds on the TURs for the steady-state particle and heat currents under a strong constraint on the Onsager coefficients.

## II. MODEL AND THEORY

We study the TUR for a multi-terminal system within the linear response regime, focusing on the case where the time-reversal symmetry is broken by an applied magnetic field $\mathbf{B}$. The steady-state transport can be effectively analyzed using the multi-terminal model depicted in Fig.1. In this setup, the central circle represents a simple conductor that is in thermal and electrical contact with two reservoirs, which permit particles and heat exchange between the left reservoir at temperature $T_L$ and chemical potential $\mu_L$ and the right reservoir at temperature $T_R$ and chemical potential $\mu_R$. The $N-2$ terminals were originally proposed by Büttiker [60] act as a probe, which plays a crucial role and has become a widely used tool to simulate inelastic events in an otherwise conservative system. Its temperature $T_P^\alpha$ and chemical potential $\mu_P^\alpha$ ($\alpha = 1,2,3\cdots N-2$) are adjusted to ensure that there is no net exchange of particles or heat with the conductor. This is a crucial additional condition for deriving Eq. (5) below (for specific details, see Ref. [61]). We choose the reference values for temperature and chemical potential to be $T_R = T$ and $\mu_R = \mu$. The temperature and chemical potential of the left reservoir and probe terminals are then defined as $T_L = T + \Delta T$, $\mu_L = \mu + \Delta\mu$ and $T_P^\alpha = T + \Delta T_P^\alpha$, $\mu_P^\alpha = \mu + \Delta\mu_P^\alpha$, respectively. In order to maintain the model as simple as possible, we consider noninteracting electrons, which means that no inelastic scattering events occur outside the reservoirs and that electron transfer between them remains coherent. We assume that both the temperature difference $\Delta T$ and $\Delta T_P^\alpha$ and the chemical potential difference $\Delta\mu$ and $\Delta\mu_P^\alpha$ are small compared to the respective reference values, ensuring that the system remains within the framework of linear irreversible thermodynamics. Once the system reaches a steady state, the steady-state particle current $J_\rho$ and the heat current $J_q$ flow from the left reservoir to the right reservoir, driven by the thermodynamic forces $X_\rho = \Delta\mu/T$ and $X_q = \Delta T/T^2$ .

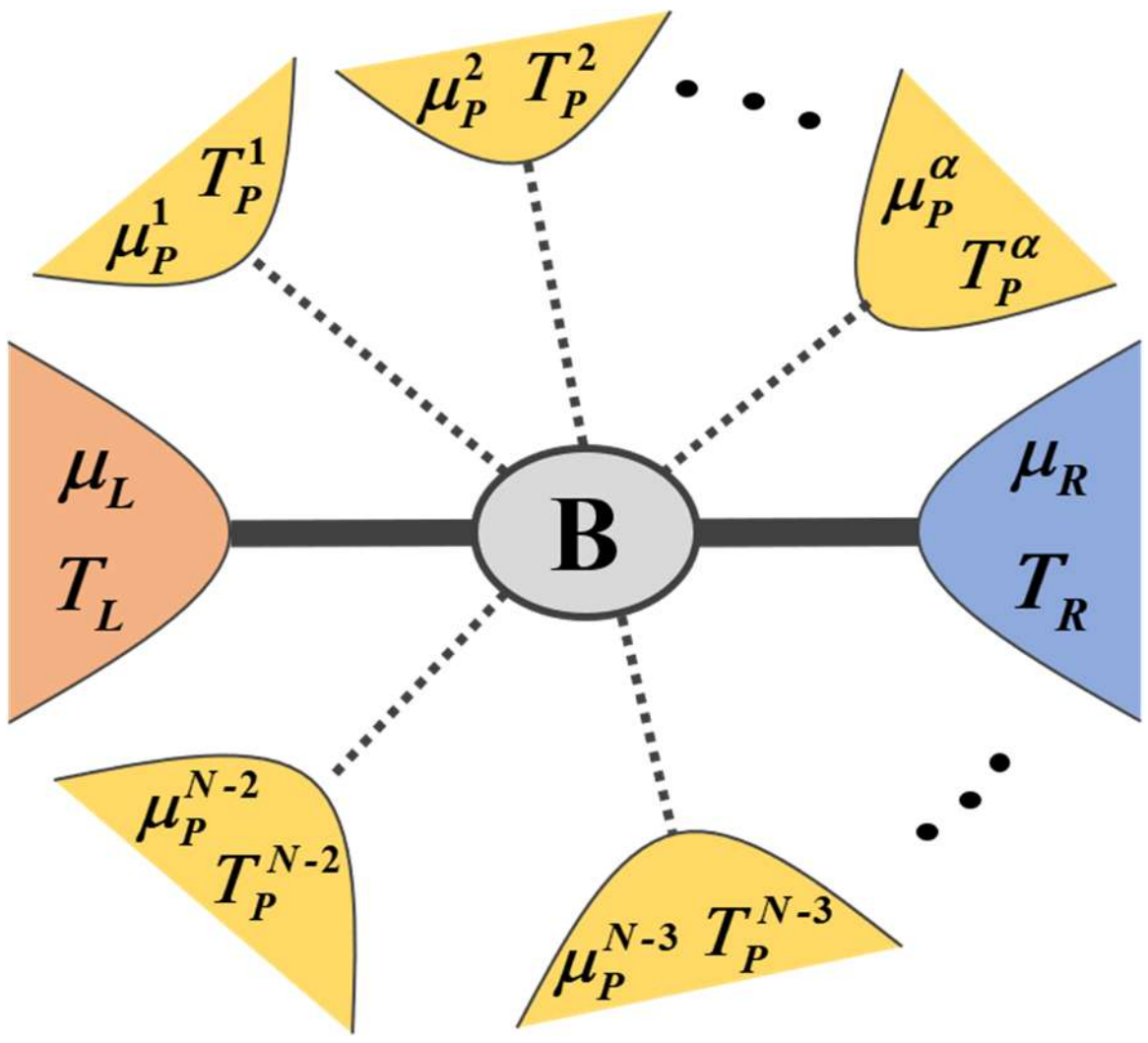

Fig. 1. The schematic diagram of a multi-terminal system, where a simple conductor is connected to $N$ independent reservoirs with different temperatures and chemical potentials in the presence of a magnetic field **B**. The steady-state transport is achieved between two transport terminals ( $L$ and $R$ ), while the other $N-2$ terminals serve as probe terminals, with their temperature and chemical potential adjusted to ensure no net exchange of particles or heat with the conductor.

According to the linear irreversible thermodynamics, the relationship between the currents and the thermodynamic forces are described by the following the phenomenological equations

$$\begin{pmatrix} J_\rho \\ J_q \end{pmatrix} = \begin{pmatrix} L_{\rho\rho} & L_{\rho q} \\ L_{q\rho} & L_{qq} \end{pmatrix} \begin{pmatrix} X_\rho \\ X_q \end{pmatrix}. \tag{2}$$

These relations are referred to as coupled transport equations. The coefficients $L_{ij}$ ( $i,\ j = \rho, q$ ) are known as the Onsager coefficients, which satisfies the Onsager-Casimir relation in the presence of a magnetic field, i.e., $L_{\rho q}(\mathbf{B}) = L_{q\rho}(-\mathbf{B})$ . The total entropy rate can be written in the linear combination of currents and the corresponding thermodynamic forces as

$$\begin{aligned}\sigma &= J_\rho X_\rho + J_q X_q \\ &= L_{\rho\rho} X_\rho^2 + L_{qq} X_q^2 + \left(L_{\rho q} + L_{q\rho}\right) X_\rho X_q\end{aligned}. \tag{3}$$

However, in a noninteracting multi-terminal system, an additional constraint on the Onsager coefficients arises from current conservation, i.e., so-called $N-2$ probe terminals, whose temperature and chemical potential are adjusted such that they do not contribute to the net charge or energy transport. This results in a stricter constraint on the Onsager coefficients [61] i.e.,

$$4L_{\rho\rho}L_{qq} - \left(L_{\rho q} + L_{q\rho}\right)^2 \geq \tan^2\left(\frac{\pi}{N}\right)\left(L_{\rho q} - L_{q\rho}\right)^2. \tag{4}$$

This constraint generally represents a stronger inequality than the second law of thermodynamics, which requires only $4L_{\rho\rho}L_{qq} - \left(L_{\rho q} + L_{q\rho}\right)^2 \geq 0$. It reduces to the bare second law only in the case where time-reversal symmetry $L_{\rho q} = L_{q\rho}$ is preserved.

## III. THE TUR OF THE PARTICLE CURRENT

We now first consider the TUR of the particle current, which express the quantity $Q_{J_\rho}$ defined in Eq. (1) as

$$Q_{J_\rho} = \frac{D_{J_\rho}\sigma}{{J_\rho}^2 k_B}, \tag{5}$$

where $D_{J_\rho}$ is the fluctuation of particle current, originating from the fluctuation-dissipation relation in the linear response [47, 58]

$$D_{J_\rho} = 2k_B L_{\rho\rho}. \tag{6}$$

Based on the strong constraint on the Onsager coefficients of Eq. (4), the Onsager coefficient $L_{qq}$ can be rewritten as

$$L_{qq} \geq \frac{\tan^2\left(\frac{\pi}{N}\right)\left(L_{\rho q} - L_{q\rho}\right)^2 + \left(L_{\rho q} + L_{q\rho}\right)^2}{4L_{\rho\rho}}. \tag{7}$$

When the equality sign is reached, a bound of $Q_{J_\rho}$ can be derived as

$$Q_{J_\rho}^{bound} = 2\left[1-(1-x)\frac{4(1+L)-(1-x)\left(1+\tan^2(\pi/N)\right)}{4(1+L)^2}\right], \tag{8}$$

where $x \equiv L_{q\rho}/L_{\rho q}$ is the asymmetry parameter, and $L \equiv L_{\rho\rho}X_\rho/\left(L_{\rho q}X_q\right)$ is the dimensionless parameter for particle current. The Eq. (8) provides the exact expression for the bound of $Q_{J_\rho}$. It clearly indicates that the $Q_{J_\rho}^{bound}$ is not only dependent on the number of terminals, but also strongly dependent on the asymmetric parameter $x$ (regulated by the external magnetic field) and system parameter $L$. This is one of the main conclusions of this paper.

In Fig. 2, we present numerical results for the bound of TUR for the particle current $Q_{J_\rho}^{bound}$ based on Eq. (8). Specifically, Fig. 2(a) illustrates $Q_{J_\rho}^{bound}$ as a function of the dimensionless parameter $L$ for different $N$. It is observed that $Q_{J_\rho}^{bound}$ violates the previously established lower bound $Q_J \geq 2$, which was originally derived based on Markovian dynamics for two-terminal systems. Eq. (8) returns to $Q_{J_\rho}^{bound} = 2$ only in the presence of the time reversal symmetry ($x=1$) within the linear response regime. We further verify this in Fig. 2 (b), where $Q_{J_\rho}^{bound}$ is plotted as a function of $x$ for different $N$. Interestingly, as shown in Fig. 2, the curves of $Q_{J_\rho}^{bound}$ exhibit a minimum value. This minimum value can be obtained from Eq. (8) with respect to $L$, i.e., $\partial Q_{J_\rho}^{bound}/\partial L = 0$, in which case $L$ satisfies the relation

$$L = \frac{1-x}{2}\tan^2\left(\frac{\pi}{N}\right) - \frac{1+x}{2}. \tag{9}$$

As a result, the minimum value of $Q_{J_\rho}^{bound}$ is given by

$$\min Q_{J_\rho}^{bound} = 2\sin^2\left(\frac{\pi}{N}\right). \tag{10}$$

This result is in exact agreement with that reported in Ref. [54]. It should be emphasized that this outcome merely indicates that the minimum value of $Q_{J_\rho}^{bound}$ depends solely on the number of terminals. For example, $\min Q_{J_\rho}^{bound}=1.5$ for $N=3$, $\min Q_{J_\rho}^{bound}=1$ for $N=4$, and $\min Q_{J_\rho}^{bound}=0.5$ for $N=6$, as shown in Fig . 2. However, the achievement of this minimum value occurs only under a specific set of system parameter conditions, as defined by Eq. (9), which depends not only on the number of terminals but also on the specific values of the system parameters.

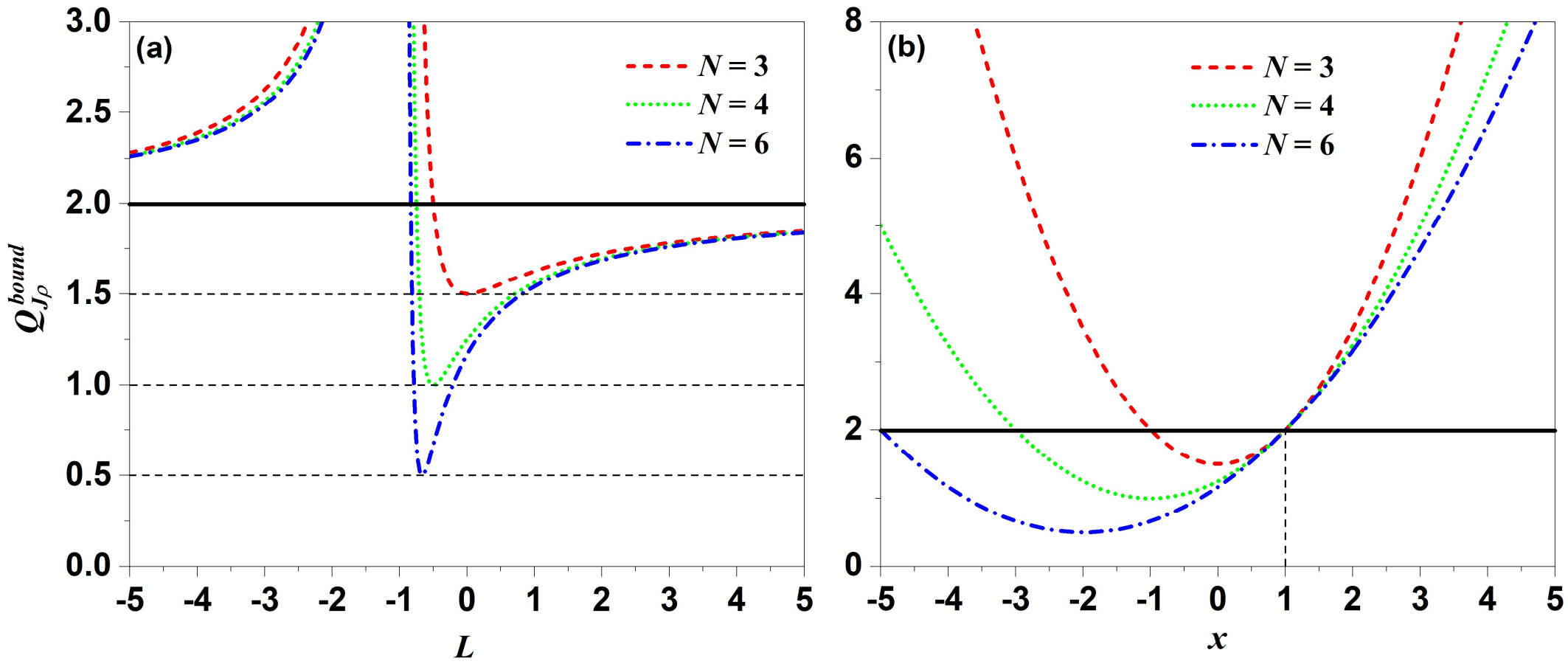

Fig. 2. The bound of the TUR for the particle current $Q_{J_\rho}^{bound}$ as a function of the dimensionless parameter $L$ (a) and the asymmetry parameter $x$ (b) for different $N$. The solid black lines represent the lower bound $Q_{J_\rho}^{bound}=2$ for time-reversal symmetry ($x=1$) in the linear response regime. The parameters $x=0.5$ for (a), and $L=1$ for (b).

## IV. THE TUR OF THE HEAT CURRENT

Analogous to the case of the particle current, the TUR of the heat current can be expressed as

$$Q_{J_q}=\frac{D_{J_q}\sigma}{J_q^{\,2}k_B}. \tag{11}$$

The fluctuation of heat current $D_{J_q}$ in linear response can be written as [47, 58]

$$D_{J_q}=2k_B L_{qq}. \tag{12}$$

Based on the inequality derived from Eq. (4), the Onsager coefficient $L_{\rho\rho}$ reads

$$L_{\rho\rho} \geq \frac{\tan^2\left(\frac{\pi}{N}\right)\left(L_{\rho q} - L_{q\rho}\right)^2 + \left(L_{\rho q} + L_{q\rho}\right)^2}{4L_{qq}}, \tag{13}$$

then we can derive the bound of $Q_{J_q}$ as

$$Q_{J_q}^{bound} = 2\left[1 + (1-x)\frac{4xM + 3x + 1 + (1-x)\tan^2(\pi/N)}{4(x + xM)^2}\right], \tag{14}$$

where $M \equiv L_{qq}X_q / \left(L_{q\rho}X_\rho\right)$ is the dimensionless parameter for heat current. In the presence of time-reversal symmetry, i.e., $x = 1$, Eq. (14) is simplified to $Q_{J_q}^{bound} = 2$. By comparing with Eq. (8), it can be observed that the bound of $Q_{J_q}$ is different from that of the particle current $Q_{J_\rho}$. The curves of $Q_{J_q}^{bound}$ as a function of $M$ and $x$ for different $N$ are shown in Fig. 3(a) and 3(b), respectively. Obviously, it can be seen that $Q_{J_q}^{bound}$ also possesses a minimum value, when the dimensionless parameter

$$M = \frac{(x-1)\tan^2(\pi/N) - (x+1)}{2x}, \tag{15}$$

Therefore, the minimum value of $Q_{J_q}^{bound}$ can be given by

$$\min Q_{J_q}^{bound} = 2\sin^2\left(\frac{\pi}{N}\right). \tag{16}$$

This is consistent with the form of the minimum value of $Q_{J_\rho}^{bound}$. By comparing Eqs. (9) and (15), it can be observed that although the $Q_{J_\rho}^{bound}$ and the $Q_{J_q}^{bound}$ exhibit the same mathematical form for their respective minimum values, they cannot simultaneously attain these minima due to differing conditions required for minimization. This is also one of the main conclusions of this paper.

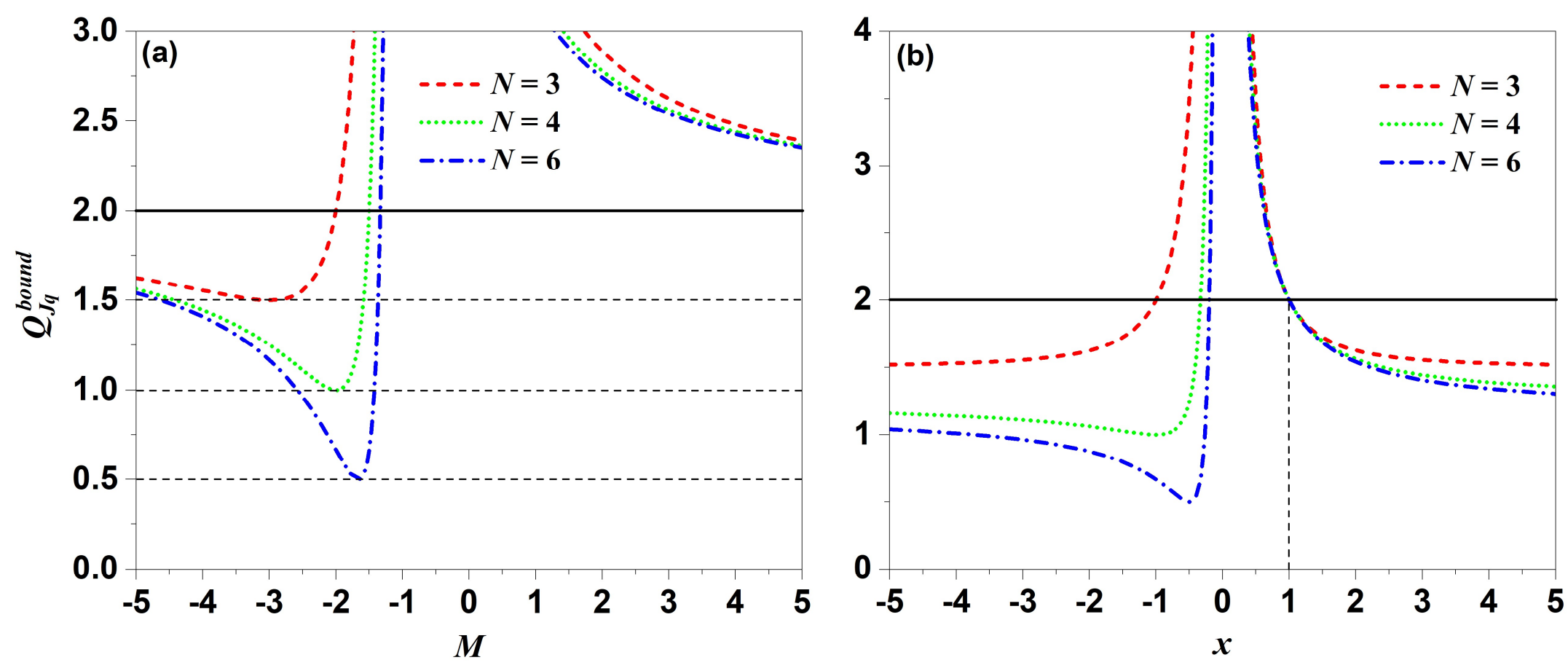

Fig. 3. The bound of the TUR for heat current $Q_{J_q}^{bound}$ as a function of the dimensionless parameter $M$ (a) and the asymmetry parameter $x$ (b) for different $N$. The parameters $x = 0.5$ for (a), and $M = 1$ for (b).

## V. CONCLUSIONS

We have investigated the TUR in the context of steady-state transport in multi-terminal systems with broken time-reversal symmetry. We independently derived exact bounds for both particle and heat currents under a stringent constraint imposed on the linear transport Onsager coefficients. We have found that the bounds of the particle and heat currents are different, and they are not only dependent on the number of terminals, but also strongly dependent on the external magnetic field and system parameters. In addition, we have determined that under specific parameter conditions, the TURs of the particle and heat currents have a unified minimum value, which only depends on the number of terminals. However, although the $Q_{J_\rho}^{bound}$ and the $Q_{J_q}^{bound}$ exhibit the same mathematical form for their respective minimum values, they cannot simultaneously attain these minima due to differing conditions required for minimization. These findings highlight the close relationship between thermodynamic uncertainty relations and the configuration of the system, which may stimulate greater research interest.

## ACKNOWLEDGEMENTS

This paper is supported by the National Natural Science Foundation of China (No. 12365006 and No. 12304058), the Natural Science Foundation of Guangxi (China) (No. 2022GXNSFBA035636), Fujian Provincial Natural Science Foundation (No. 2023J01006), Fundamental Research Funds for Central Universities of the Central South University (No. 20720240145).